\documentclass{edp-conf}
\usepackage{graphicx}

\begin{document}

\TitreGlobal{SF2A 2005}

%%-----------------------------
%%      the top matter
%%-----------------------------
\title{Near-UV to near-IR disk-averaged Earth's spectra from Moon's Earthshine observations}
\author{Hamdani, S.}\address{Observatoire de Haute Provence (CNRS-OAMP), 04870 St Michel l'Observatoire, France}
\author{Arnold, L.$^1$}
\author{Foellmi, C.}\address{ESO, Casilla 19001, Santiago 19, Chile}
\author{Berthier, J.}\address{IMCCE, Observatoire de Paris, 77 Avenue Denfert-Rochereau, 75014 Paris, France}
\author{Briot, D.}\address{Observatoire de Paris-Meudon, 5 place Jules Janssen 92195 Meudon, France}
\author{Francois, P.$^4$}
\author{Riaud, P.}\address{Institut d'Astrophysique et de G\'eophysique de Li\`ege, Universit\'e de Li\`ege, All\'ee du 6 Ao\^ut, 4000, Sart-Tilman, Belgium}
\author{Schneider, J.$^4$}

\runningtitle{Hamdani, Arnold \etal}
\setcounter{page}{237}
% Keep this line, even if the page will be settled afterwards..
\index{Author1, A.}
\index{Author2, B.}
\index{Author3, C.}
% Repeat the authors here, this will help to make the final index

\maketitle
\begin{abstract}
We discuss a series of Earthshine spectra obtained with the NTT/EMMI instrument between 320nm and 1020nm with a resolution of R$\cong$450 in the blue and R$\cong$250 in the red. These ascending and descending Moon's Earthshine spectra taken from Chile give disk-averaged spectra for two different Earth's phases. The spectra show the ozone (Huggins and Chappuis bands), oxygen and water vapour absorption bands, and also the stronger Rayleigh scattering in the blue. Removing the known telluric absorptions reveals a spectral feature around 700nm which is attributed to the vegetation stronger reflectivity in the near-IR, so-called vegetation red-edge.
\end{abstract}

\section{Introduction}
Since the first measurements of Earth disk-averaged reflectance spectra by Arnold et al. (2002) and Woolf et al. (2002), several attempts have been successful in the same spectral bandwidth (\cite{seager05}, \cite{montanes05}). Most of these spectra show signatures of Earth atmosphere and ground vegetation.

\section{Earth's reflectance spectra}
We get four Earthshine Moon's spectra at the NTT/EMMI telescope in ESO/Chile during the nights of the 09-18-04, 05-24-04 and more recently, 05-31-05 and 06-02-05. The spectra cover the domain from 320 to 1020nm with a gap of 20nm between 510 and 530nm. The resolution is  R$\approx$450 in the blue and R$\approx$250 in the red. We extracted the Earth's integrated reflectance spectra by correcting these spectra from the Moon reflectance, the Sun spectra and the atmosphere transmittance. We also take into account the changing colour of the integrated Moon versus its observation phase (\cite{lane73}).

\begin{figure}[h]
   \centering
  \includegraphics[width=13cm]{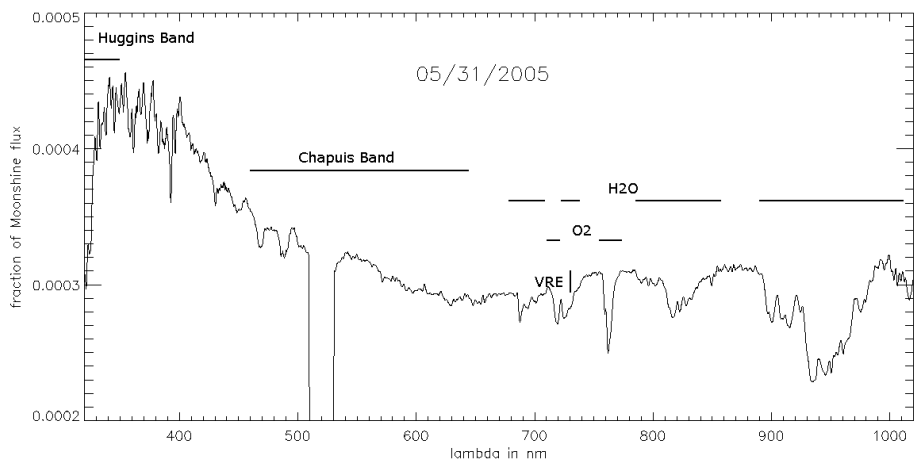}
      \caption{Earth relfectance spectra for the West Africa and part of the Atlantic Ocean.}
       \label{spectre}
   \end{figure}

We can identify on each reflectance spectrum the absorption bands of $O_2$, $O_3$ and $H_2 O$. The Rayleigh scattering is well visible, as well a the strong absorption of ozone Huggins band from 320nm to 350nm (Fig. \ref{spectre}).

It is known that the vegetation has a stronger reflectivity in the near IR above 700nm with respect to the visible. To measure it, we correct our reflectance spectra from the atmospheric absorption lines and the Rayleigh scattering to get Earth's ground reflectance. By comparing the flux in two spectral domains bracketing the 700nm rise, we measure the Vegetation Red-Edge (VRE) relative to the continuum. We measure VRE$\approx10\%$ when the Earth facing the Moon shows Africa and Europe, and VRE$=3\%$ with the Pacific Ocean and a small part of North America. This demonstrates how Earth's phase influences the detectivity of the vegetation.

\end{document}